\documentclass[aps,prl,preprint,superscriptaddress]{revtex4}
 
\usepackage{bm}

\begin{document}

\title{Photonuclear Sum Rules and the Tetrahedral Configuration
of $^4$He}
\author{Doron Gazit}
\affiliation{The Racah Institute of Physics, The Hebrew University, 
     91904 Jerusalem, Israel}

\author{Nir Barnea}
\affiliation{The Racah Institute of Physics, The Hebrew University, 
     91904 Jerusalem, Israel}

\author{Sonia Bacca}
\affiliation{Gesellschaft f\"ur Schwerionenforschung,
Planckstr.~1, 64291 Darmstadt, Germany}

\author{Winfried Leidemann}
\affiliation{Department of Physics, George Washington University, Washington DC 20052, USA,\\
and Istituto Nazionale di Fisica Nucleare, Gruppo Collegato di Trento}
\altaffiliation{On leave of absence from Department of Physics, University of Trento, I-38050 Povo (Trento) Italy}

\author{Giuseppina Orlandini}\affiliation{Department of Physics, George Washington University, Washington DC 20052, USA,\\
and Istituto Nazionale di Fisica Nucleare, Gruppo Collegato di Trento}
\altaffiliation{On leave of absence from Department of Physics, University of Trento, I-38050 Povo (Trento) Italy}

\date{\today}

\begin{abstract}
Three well known photonuclear sum rules (SR), i.e. the Thomas-Reiche-Kuhn, 
the bremsstrahlungs and the polarizability SR are calculated for $^4$He
with the realistic nucleon-nucleon potential Argonne V18 and the 
three-nucleon force Urbana IX. The relation between these sum rules and the corresponding 
energy weighted integrals of the cross section is discussed. Two additional equivalences 
for the bremsstrahlungs SR are given, which connect it to the proton-neutron 
and neutron-neutron distances. Using them, together with our result
for the bremsstrahlungs SR, we find a deviation from the tetrahedral symmetry
of the spatial configuration of $^4$He. The possibility to access this deviation experimentally is discussed.
\end{abstract}

\maketitle

Photonuclear sum rules (SR) are related to moments of different order of the 
photonuclear cross section and reflect important electromagnetic properties of nuclei.
In fact they can often be expressed in terms of simple ground state properties in a model 
independent or quasi model independent way. 
Well known examples are the Thomas-Reiche-Kuhn (TRK) sum rule~\cite{TRK}, which 
gives information about the importance of exchange effects in nuclear 
dynamics via the so-called TRK enhancement factor $\kappa^{\rm{TRK}}$, the bremsstrahlungs 
sum rule (BSR) ~\cite{LevingerBethe,Brink,Foldy,DellafioreBrink}, 
which is connected to the nuclear charge 
radius and to the mean distance between protons~\cite{DFlippa}, 
and the polarizability  
sum rule (PSR)~\cite{Friar},  related to the electric nuclear polarizability. 
These SR are all assuming that the dominant contribution to the cross section comes 
from unretarded electric dipole (E1UR) transitions. Two- and three-body studies~\cite{SaA,Benchmark} indeed
confirm that other contributions are much smaller. 
Much discussed is also the Gerasimov-Drell-Hearn (GDH) sum rule~\cite{GDH},
which is related to the nuclear anomalous magnetic moment. 

In this work we consider the  TRK, BSR and PSR of $^4$He within a 
realistic nuclear potential model consisting of two- and  three-body forces
(AV18 and UIX~\cite{AV18,UIX}). (The GDH sum rule is trivial for $^4$He: it vanishes, 
since the $^4$He total angular momentum is equal to zero.)
We also investigate the related moments by integrating 
explicitly the properly weighted total photoabsorption cross
section, which we have calculated  for the same potential model~\cite{prl06}.

The aim of this study is to show that in some cases sum rules can allow to
access experimentally two-body properties of the nuclear ground state, like
the proton-proton, neutron-neutron and  proton-neutron distances. In the case of $^4$He this
allows to test the validity of the configuration tetrahedral symmetry
of this nucleus and at the same time "measuring" the amount of symmetry breaking.
This work aims also at providing a guideline for 
experiments, where only lower bounds for the SR
can be determined,  as well as at giving an idea of the reliability of the SR 
approach to heavier systems, where the direct theoretical determination of the cross section, and therefore its integration, is presently out of reach. 
The advantage to perform this kind of study in $^4$He, 
compared to analogous ones in the two-~\cite{2bodySR} and 
three-body systems~\cite{ELO97pol}, is that $^4$He is a rather dense 
and compact nucleus, resembling heavier systems more closely.  
Only now that realistic theoretical results for 
the photonuclear cross 
section are available such a study is possible and one  can put the 
extrapolation of the results to heavier systems on safer grounds.

We start by recalling the formalism of the photonuclear SR. The various  moments of the 
photonuclear cross section are defined as 
\begin{equation}
m_n(\bar\omega) \,\,\, \equiv \int_{\omega_{th}}^{\bar\omega}\,d\omega \,\omega^n\,
\sigma_\gamma^{\rm E1UR}(\omega)\,,\label{moments} 
\end{equation}
where $\omega$ is the photon energy and $\omega_{th}$ and $\bar\omega$ indicate threshold energy and upper integration limit, respectively.  
With $\sigma_\gamma^{\rm E1UR}(\omega)$ we indicate 
the unretarded dipole cross section given by 
\begin{equation}
\sigma_\gamma^{\rm E1UR}(\omega)={\cal G}\, \omega\, R(\omega)\,,\label{sigma} 
\end{equation}
where ${\cal G}= 4\pi^2 \alpha/(3(2 J_0+1))$ with $\alpha$ and $J_0$ denoting the fine
structure constant and the nucleus total spin, respectively.
The dipole response function $R(\omega)$ is given by 
\begin{equation}
R(\omega) = \sum_n |\langle n| {\bf D}| 0\rangle|^2\,
            \delta(\omega-E_n+E_0) \,,
\end{equation}
where $|0/n\rangle$ and $E_{0/n}$ are the nuclear ground/excited state wave functions and energies,
respectively and ${\bf D}$ is the unretarded dipole operator ${\bf D}=\sum_{i=1}^A {\bf r}_i \tau_i^{3}/2$, where $A$ is the number of nucleons and $\tau^3_i$ and ${\bf r}_i$ are the third component of the isospin 
operator and the coordinate of the $i$th particle in the center of mass frame, respectively.

Assuming that $\sigma_\gamma^{\rm E1UR}(\omega)$ converges to zero 
faster than $\omega^{-n-1}$ and applying the closure property of the  eigenstates of 
the hamiltonian $H$, one has the following SR for $n=0,-1,-2$:
\begin{eqnarray}
\Sigma^{\rm TRK}&\equiv&m_0(\infty)\,
 = \,\frac{{\cal G}}{2}\,
\langle 0| \left[{\bf D}  ,\left[ H, {\bf D} \right]\right]|0\rangle
\label{mTRK}\\
\Sigma^{\rm BSR}&\equiv& m_{-1}(\infty)\, 
 = \,{\cal G}\,\langle 0|  {\bf D} \cdot  {\bf D}|0
\rangle \,
\label{mBSR} \\
\Sigma^{\rm PSR}&\equiv& m_{-2}(\infty)\,
 = \,{\cal G}\,\sum_n (E_n-E_0)^{-1}|\langle n|{\bf D}|0\rangle|^2\,,\label{mPSR}
\end{eqnarray}

Working out the expressions in Eqs.~(\ref{mTRK}-\ref{mPSR}), one finds that those moments are 
related to interesting properties of the system under consideration.
In fact the TRK sum rule is also given by the well known relation~\cite{TRK}
\begin{equation}
\Sigma^{\rm TRK}\,=\,{\cal G} \frac{3NZ}{2mA}
\left(1+\kappa^{TRK}\right),\label{TRK}
\end{equation}
where $N$ and $Z$ are the neutron and proton numbers, respectively, $m$ is the nucleon mass
and $\kappa^{TRK}$ is the so-called TRK enhancement factor defined as 
\begin{equation}
\kappa^{TRK} \equiv \frac{mA}{3NZ} 
\langle 0|[ {\bf D},[ V, {\bf D}]]|0\rangle\,.
\label{kappa}
\end{equation}
From this expression
it is evident that $\kappa^{TRK}$  embodies the  exchange effects of the nuclear potential $V$
(the double commutator in (\ref{kappa}) vanishes for systems like atoms, where no exchange 
effects are present).

In the literature one finds a few interesting equivalences for the bremsstrahlungs sum rule. 
Rewriting the dipole operator as ${\bf D} = (NZ/A){\bf R}_{PN}$, where ${\bf R}_{PN}$ denotes 
the distance between the proton and neutron centers of mass, one has~\cite{Brink}
\begin{equation}
\Sigma^{\rm BSR}\,={\cal G}\,\left(\frac{N Z}{A}\right)^2\langle 0|R_{PN}^2|0\rangle\,.
\label{BSRBrink}
\end{equation}
In Ref.~\cite{Foldy} Foldy demonstrated that
\begin{equation}
\Sigma^{\rm BSR}\,={\cal G}\,\frac{N  Z }{A-1}\langle r_p^2 \rangle\,,
\label{BSRFoldy}
\end{equation}
where $ \langle r_p^2 \rangle$ is the mean square (m.s.) point proton radius 
\begin{equation}
\langle r_p^2 \rangle\equiv \frac{1}{Z}\langle 0|\sum_{i=1}^Z r_i^2 |0\rangle\,.
\label{rpsquare}
\end{equation}
However, this relation is valid only under the assumption that the ground state wave function is symmetric in the space coordinates of the nucleons.

In Ref.~\cite{DellafioreBrink} it was found that, in the framework of 
the oscillator shell model, one has
\begin{equation}
\Sigma^{\rm BSR}\,={\cal G}\,\left(Z^2\langle r_p^2
\rangle - Z\langle r_p^{'2}\rangle\right)\,, 
\label{BSRDellafioreBrink}
\end{equation}
where $\langle r_p^{'2}\rangle$ is the m.s. distance of protons with respect to the proton 
center of mass ${\bf R}_P$
\begin{equation}
\langle r_p^{'2} \rangle\equiv \frac{1}{Z}\langle 0|\sum_{i=1}^Z ({\bf r}_i-{\bf R}_P)^2 |0\rangle\,.
\label{rpprimesquare}
\end{equation}
Later, in  Ref.~\cite{DFlippa}, it was shown that the validity of Eq.~(\ref{BSRDellafioreBrink}) is not
limited to the oscillator shell model, but it is a model independent relation, which can also be 
written as 
\begin{equation}
\Sigma^{\rm BSR}\,={\cal G}\,\left(Z^2\langle r_p^2
\rangle - \frac{Z(Z-1)}{2}\langle r_{pp}^2
\rangle\right)\,,
\label{BSRDFLippa}
\end{equation}
where $\langle r_{pp}^2\rangle$ is the m.s. proton-proton distance
\begin{equation}
\langle r_{pp}^2 \rangle \equiv \frac{1}{Z(Z-1)}\langle 0|\sum_{i,j=1}^{Z} 
({\bf r}_i- {\bf r}_j)^2|0\rangle\,.\label{r_pp}
\end{equation}

For the BSR two additional relations exist, which are easy to prove, but which,
to our knowledge, have not been considered in the literature, i.e.
\begin{equation} 
\Sigma^{\rm BSR}\,  =  \,{\cal G}\,\left(N^2\langle r_n^2
\rangle - \frac{N(N-1)}{2}\langle r_{nn}^2
\rangle\right)\, \label{BSR3} 
\end{equation}
and
\begin{equation} 
\Sigma^{\rm BSR}\,  =  {\cal G}\, \frac{NZ}{2}
\left(
\langle r_{pn}^2 \rangle\,-\, \langle r_p^2 \rangle \,-\, \langle r_{n}^2\rangle
\right)
\,,\label{BSR4} 
\end{equation}
where $\langle r_n^2\rangle$ is the m.s. point neutron radius and 
$\langle r_{\alpha\beta}^2 \rangle$ are the m.s. nucleon-nucleon (NN) distances, i.e.
\begin{eqnarray}
\langle r_{nn}^2 \rangle &\equiv& \frac{1}{N(N-1)}\langle 0|\sum_{i,j=1}^{N} 
({\bf r}_i- {\bf r}_j)^2|0\rangle\,,\label{r_nn}\\
\langle r_{pn}^2 \rangle &\equiv& \frac{1}{NZ}\langle 0|\sum_{i=1}^Z \sum_{j=1}^N
({\bf r}_i- {\bf r}_j)^2|0\rangle\,.\label{r_pn}
\end {eqnarray}
It is interesting to note that Eqs.~(\ref{BSRDFLippa}),~(\ref{BSR3}) and~(\ref{BSR4})
express $\Sigma^{\rm BSR}$ via a one-body ($\langle r_\alpha^2 \rangle$) as well as  a two-body
quantity ($\langle r_{\alpha\beta}^2\rangle$).

Finally, regarding the polarizability sum rule, one has 
\begin{equation}
\Sigma^{BSR}= 2 \pi^2 \alpha_D\,,
\end{equation} 
where $\alpha_D$ denotes the 
nuclear polarizability in the E1UR approximation. 

In this work we calculate moments and sum rules in different ways.
On the one hand we obtain the moments by integrating our recent result for the 
$^4$He total photoabsorption cross section~\cite{prl06}. This is an {\it ab initio}
calculation where we use the AV18 NN potential and the UIX 
three-nucleon force. The results have been obtained by
the Lorentz integral transform (LIT) method \cite{ELO94}. The necessary equations 
have been  solved  via hyperspherical harmonics expansions (EIHH approach~\cite{EIHH,EIHH_3NF}). 
On the other hand we obtain the SR in a more direct way, as explained in the following. 

The LIT, an integral transform with a Lorentzian kernel, is defined as follows
\begin{equation}
{\cal L}(\epsilon,\Gamma) = \int d\omega {\frac {R(\omega)} {(\omega-\epsilon)^2 
                              + \Gamma^2} } \,.
\end{equation}
One way to evaluate the LIT is by using the Lanczos technique as described in 
Ref.~\cite{Mario}. In fact the LIT can be re-expressed as
\begin{equation} \label{loreins}
{\cal  L}(\epsilon, \Gamma)=\frac{1}{\Gamma}\,
            \langle 0 | {\bf D}\cdot {\bf D}|0\rangle\,
            \mbox{Im} \,\{\langle \phi_0 |
                      \frac{1}{z-H}|\phi_0\rangle\} \mbox{} 
\end{equation}
with $z=E_0+\epsilon+ i\Gamma$ and
\begin{equation} \label{start}
  |\phi_0\rangle =\frac{{\bf D}|0\rangle }{\sqrt{\langle
  0|{\bf D}\cdot {\bf D}|0 \rangle }} \,.
\end{equation}
It is evident that the LIT depends on the matrix element
\begin{equation}
  x_{00}(z)= \langle \phi_0 |\frac{1}{z-H}|\phi_0\rangle \,,
\end{equation}
which can be expressed~\cite{Dagotto} as a continued fraction containing the Lanczos coefficients
\begin{equation}
   a_i=\langle \phi_i |H|\phi_i \rangle \mbox{,}\,\,\,\,\,\,
   b_i=\parallel b_i|\phi_i \rangle \parallel \mbox{,}
\end{equation}
where the $\phi_i$ form the Lanczos orthonormal basis $\{|\phi_i
\rangle \mbox{,}i=0,\ldots \mbox{,} n\}$.
Therefore the implementation of the Lanczos algorithm leads to ${\cal L}(\epsilon,\Gamma)$ 
(for details see ~\cite{Mario}).
While the inversion of the LIT \cite{inversion} gives access to $R(\omega)$, 
and thus to the moments of Eq.~(\ref{moments}), the normalization of the Lanczos "pivot" 
$|LP\rangle= {\bf D}|0 \rangle$ and the Lanczos coefficients allow to obtain the SR of
Eqs.~(\ref{mTRK}-\ref{mPSR}). 
In fact one has:
\begin{eqnarray} 
\Sigma^{\rm PSR} & = & {\cal G}\,x_{00} (E_0)\,,\label{LPSR}\\ 
\Sigma^{\rm BSR} & = & {\cal G}\,\langle LP |LP\rangle\,,\label{LBSR}\\ 
\Sigma^{\rm TRK} & = & (a_0- E_0)\,\Sigma^{\rm BSR} \,.\label{LTRK}
\end{eqnarray}
We use a HH basis, therefore the ground state $|0\rangle$, the Lanczos "pivot" $|LP\rangle$
and the Lanczos coefficients $a_n$
are given in terms of HH expansions.  While for the ground state the expansion is 
characterized by an even hyperspherical grand-angular quantum number $K$ and total 
isospin T=0, $|LP\rangle$ has to be expanded on $\{K'=K+1, T=1\}$  states
(we neglect the AV18 isospin mixing, which is very small, as shown for the $^4$He ground 
state in~\cite{pisaIM}).
The rate of convergence of the various SR results from Eqs.(\ref{LPSR}-\ref{LTRK}) 
is given in Table~\ref{Table1} as a function of the hyperspherical grand-angular quantum number $K$. 
One observes sufficiently good convergence patterns for $\Sigma^{\rm TRK}$ and $\Sigma^{\rm BSR}$.
For the latter we have performed an additional test of the convergence, calculating 
$\Sigma^{\rm BSR}$ directly as mean value of the operator ${\bf D} \cdot {\bf D}$ on the ground state 
(Eq.~(\ref{mBSR})). In this way the expansion of $|LP\rangle$ on $\{K'=K+1, T=1\}$ states is avoided. 
We obtain practically identical results.

From Table~\ref{Table1} one sees that the convergence of $\Sigma^{\rm PSR} $ is slower when the three-nucleon force (3NF) is included. In~\cite{prl06} a related problem has been found for the cross section itself. 
In fact it has been shown
that the peak of the giant dipole resonance is slightly shifting towards lower energies with 
increasing $K$. 
The sum rule $\Sigma^{\rm PSR}$, which has the strongest inverse 
energy weighting, is more sensitive to this shift than the other two SR.
In~\cite{prl06} an extrapolated cross section $\sigma_\gamma^{\infty}$ was obtained from 
a Pad\`e approximation. We have used this to determine from Eq.~(\ref{moments})
the various moments for $\bar\omega=$ 300 and 135 MeV.
These results are also listed in Table~\ref{Table1}. 

For $\Sigma^{\rm TRK}$ one sees that the SR
is not yet exhausted at 300 MeV. In fact more than 20\% of the strength is still missing.
At pion threshold, one has only about 2/3 of $\Sigma^{\rm TRK}$. 
As was already discussed in~\cite{ELOT00} for the triton case, the rather 
strong contribution from higher energies seems to be connected
to the strong short range repulsion of the AV18 potential. 
As to the TRK enhancement factor we obtain $\kappa^{\rm TRK}=1.31$ for AV18 and  1.44 for AV18+UIX.
These numbers are somewhat larger than older results obtained either with a variational wave function
and AV14+UVII potential ($\kappa^{\rm TRK}=1.29$ ~\cite{Schiavilla}) or with more approximated wave functions and
various soft and hard core NN potentials ($\kappa^{\rm TRK}=0.9 - 1.30$ ~\cite{Weng,Horlacher,Gari}).

For $\Sigma^{\rm BSR}$, as expected, the contribution at high energy is much smaller 
than for $\Sigma^{TRK}$, in fact at $\bar\omega= 300$ MeV we find a missing sum rule 
strength of less than 2\% only. 
For $\Sigma^{\rm PSR}$ the strength beyond 300 MeV is even more negligible. Actually
in this case the explicit integration leads to an even higher result than the sum 
rule evaluation of Eq.~(\ref{LBSR}). The seeming contradiction is explained by 
the already discussed fact  that $\Sigma^{\rm PSR} $ is not yet convergent, 
while for the explicit integration an  extrapolated cross section 
is used. Indeed, integrating the K=18 cross section we obtain a
value of 6.46 mb, which is consistent with the corresponding sum rule
result of 6.47 mb, within the numerical error of the calculation.
For the AV18+UIX force the value of the polarizability
$\alpha_D$ that we deduce from the extrapolated $\Sigma^{\rm PSR}$ is $0.0655$ fm${^3}$.
The AV18 result, which already shows a good convergence for $K=16$, 
is  0.0768 fm${^3}$. 
This means that the 3NF reduces the polarizability by 
15\%.  
It would be very interesting to measure this nuclear polarizability by Compton 
scattering, as a test of the importance of the three-body force on such a classical 
low-energy observable.
We find  that $\Sigma^{\rm PSR}$ is the SR that is affected most by the 3NF.
In fact $\Sigma^{\rm BSR}$ is reduced by only 10\% and one has an opposite effect on 
$\Sigma^{\rm TRK }$ with a 5\% increase. 
The quenching or enhancement of SR due to the 3NF is the reflection of its
effects on the cross section i.e a decrease of the peak and an increase of the tail. 

Here we would like to add a few words about the very old question, already  
discussed in ~\cite{LevingerBethe}, of the 
"existence" of the SR (finiteness of $m_n(\bar\omega)$ for $\bar\omega\rightarrow\infty)$,
which is connected to the high-energy fall-off 
of the E1UR cross section.
Since we find a rather good consistency  between the SR and the moment values, we can
state with  a rather high degree of confidence that $\Sigma^{\rm TRK} $ (and consequently the
other two SR) "exist".
Therefore we can try to extract some information about the high energy behavior of 
the cross section and hence about the "existence" of SR with higher $n$.
With an  $\omega^{-p}$ ansatz for the fall-off of the cross section above pion threshold,  
and requiring that $\Sigma^{\rm TRK} $  is 146 mb MeV (see Table~\ref{Table1}),
one gets a rather weak energy fall-off, i.e. $ p\simeq 1.5$. This value is also consistent 
with $\Sigma^{\rm BSR}$. 
In fact, adding such a tail contribution to $m_{-1}(135)$
one gets  $\Sigma^{\rm BSR}=2.39 $ mb, to be compared with 2.41 mb in
Table~\ref{Table1}. The value of $p$ might be somewhat different for other potentials, 
but probably it will not change much. Therefore  we can rather safely conclude that
higher order SR do not exist for realistic nuclear potential models.

Finally we return to discuss the BSR, because it presents a very interesting aspect. As already mentioned $\Sigma^{\rm BSR}$  contains information about one- and two-body densities 
via $\langle r_\alpha^2 \rangle$  and $\langle r_{\alpha\beta}^2\rangle$, respectively. This means that a measurement of $\Sigma^{\rm BSR}$ and the knowledge of the experimental 
m.s. radius allow to determine $\langle r_{\alpha\beta}^2\rangle$ via Eqs.~(\ref{BSRDFLippa}),(\ref{BSR3}) and~(\ref{BSR4}). In this way one gets information about the
internal configuration of $^4$He as it is explained in the following. 

In his derivation of
Eq.~(\ref{BSRFoldy}), Foldy assumed a totally symmetric $^4$He spatial wave function, which
 corresponds to a configuration where the four nucleons are located at the four vertexes of a tetrahedron. For such a configuration one has $\langle r_p^2\rangle=\langle r_n^2\rangle=
\langle r^2\rangle$ and $\langle r_{pp}^2\rangle=\langle r_{nn}^2\rangle=
\langle r_{np}^2\rangle$  with $Q_T\equiv\langle r_{\alpha\beta}^2\rangle/\langle r^2\rangle=8/3$.
Foldy's assumption is a very good approximation for $^4$He, but other spatial symmetries
(mixed symmetry, antisymmetric) are also possible. What can be learned from our  $\Sigma^{\rm BSR}$ result with respect to this question?
For $^4$He, which is a T=0 system one can safely assume  that
$\langle r^2_p\rangle = \langle r^2\rangle= \langle r_n^2\rangle$
(isospin mixing is tiny~\cite{pisaIM}). 
Using in Eqs.~(\ref{BSRDFLippa}),~(\ref{BSR3}) and~(\ref{BSR4}) the AV18+UIX value of 
$\langle r^2\rangle = 2.04$ fm$^2$ 
~\cite{prl06} (which coincides with the experimental one~\cite{radiusexp}, 
corrected for the proton  charge radius)
and $\Sigma^{\rm BSR}$ from Table~\ref{Table1}, we obtain 
$\langle r^2_{pp}\rangle=\langle r^2_{nn}\rangle=5.67$ fm$^2$ and 
$\langle r^2_{pn}\rangle=5.34$ fm$^2$, i.e. two values which differ by about 6\%.
The ratios $Q_{pp(nn)}\equiv\langle r^2_{pp(nn)}\rangle/\langle r^2\rangle$ and 
$Q_{np}\equiv\langle r^2_{np}\rangle/\langle r^2\rangle$ are not much different from the 
correspondent value $Q_T$ of a classical tetrahedral configuration. We obtain 
$Q_{pp}=2.78$ and  $Q_{np}=2.62$ instead of  $Q_T=2.67$. One notices that $Q_{pp(nn)}-Q_T
\simeq 2 (Q_{np}-Q_T)$. This reflects the different numbers of proton-proton 
and neutron-neutron pairs (2) with respect to  proton-neutron pairs (4).
Using  Eq.~(\ref{BSRBrink}) one can also derive the distance 
between the proton and neutron centers of mass. One has $R_{pn}=$ 1.58 fm instead  of 1.65 fm for the tetrahedral configuration.

Notice that with $\langle r^2\rangle = 2.04$ fm$^2$ the "tetrahedral" BSR would be 
2.62 mb, the same value that one obtains using Eq.~(\ref{BSRFoldy}). This value is  9\% larger than our result. 
The  distortions  that we find from our 9\% smaller BSR 
are the consequence of the different effects of 
the potential on isospin triplet and isospin singlet pairs. 

We conclude that when considered in its body frame $^4$He should look 
like a slightly {\it deformed}
tetrahedron. Of course this statement has to be intended in a quantum mechanical sense, 
regarding the mean square values of the nucleon-nucleon distances on the two-body density.
It is clear that one cannot measure this {\it deformation} "directly",
since it is not a deformation of the one-body charge density (the $^4$He
charge density has only a monopole). On the other hand such a {\it deformation}
is accessible experimentally in an indirect way via the measurements of the charge radius and of
$\Sigma^{\rm BSR}$. 

This leads to the question how exactly $\Sigma^{\rm BSR}$
can be measured in a photonuclear experiment. Two points have to be addressed: i) 
the contributions of E1 retardation and higher multipoles, which are not contained in $\Sigma^{\rm BSR}$,
but which will contribute to the experimental cross section and ii) the contribution of 
the high energy tail. To this end it is instructive to
consider the results from the three-nucleon photodisintegration. Regarding point i),
additional effects of the E1 retardation and higher multipoles have been calculated 
in Ref.~\cite{Golakbenchmark} for the AV18+UIX potentials in a Faddeev 
calculation. Using those results one finds that E1 retardation 
and higher multipoles increase  $m_{-1}(135)$ by about 1\% only.
In fact according to Gerasimov\cite{Gerasimov} there is a large cancellation of E1 
retardation and other multipoles.
There is no reason for a larger effect in $^4$He. On the contrary,
since the leading isovector magnetic dipole (M1) transition is suppressed in 
this nucleus (it is zero for an S-wave) one can expect an even smaller contribution.
As to point ii)  considering the fall-off of the triton E1UR cross section around pion threshold 
from Ref.~\cite{Golakbenchmark} one obtains a result very similar to our $^4$He
case, namely $p \simeq 1.5$.  However, including the other multipole contributions
one gets a considerably smaller value, namely $p \simeq 1.1$. This is no contradiction
with the small increase of 1\% for $\Sigma^{\rm BSR}$, since the inverse energy weighted cross section integrated from 100 to 135 MeV gives only a rather small contribution to
the sum rule. On the other hand one would overestimate $\Sigma^{\rm BSR}$ taking the
full cross section. Thus we suggest that the tail contribution to an experimental 
BSR should be estimated 
using the theoretically established fall-off $\omega^{-p}$ with $p \simeq 1.5$.

In conclusion, we have studied three well known photonuclear sum rules for $^4$He within 
a realistic two- and  three-nucleon potential model. Two new equivalences for the BSR
sum rules have allowed us to deduce information about two-body properties of the nuclear ground state, like
the proton-proton, neutron-neutron and  proton-neutron distances. In particular we have tested the validity of the configuration tetrahedral symmetry
of this nucleus and have found that this symmetry is slightly broken.
We have suggested an experimental way to access this symmetry breaking via a measurement of
the BSR which could be performed in one of the existing or planned low-intermediate energy photonuclear facility.

\bigskip\bigskip
D.G, N.B, W.L and G.O. thank the Institute for Nuclear Theory at the University of 
Washington for its
hospitality and the Department of Energy for partial support during the completion of this work. The work of D.G. and N.B. was supported by the ISRAEL SCIENCE FOUNDATION
(Grant number 361/05).

\newpage
\begin{table}
\caption{
\label{Table1}
Convergence in K of the SR for AV18+UIX potentials. The converged AV18 results are
also shown. The last two lines of the table show the convergence in $\bar\omega$ of the 
various moments. 
}
\begin{center}
\begin{tabular}
{c | ccc} 
 &&  
                \,\, AV18+UIX\,\, \\ 
\hline
&$\,\Sigma^{\rm PSR}\,$ &$\,\Sigma^{\rm BSR}\,$ &$\,\Sigma^{\rm TRK}\,$\\
K &$\,10^{-2}$[mb MeV$^{-1}$]\, & \,[mb]\, & \,$10^{2}$[mb MeV]\,\\
\hline\hline
  8 &  6.230 & 2.398 & 1.430 \\ 
 10 &  6.277 & 2.396 & 1.448 \\ 
 12 &  6.331 & 2.394 & 1.451 \\ 
 14 &  6.382 & 2.401 & 1.458 \\ 
 16 &  6.434 & 2.406 & 1.460 \\ 
 18 &  6.473 & 2.410 & 1.462 \\ 
\hline\hline
&&AV18 \\
\hline &7.681 & 2.696 & 1.383 \\
\hline\hline
&& \,\, AV18+UIX\,\, \\ 
\hline
 $\bar\omega$ [MeV]&$\,m_{-2}\,(\bar\omega) \,$ & $\,m_{-1}\,(\bar\omega) \,$ & $\,m_{0}\,(\bar\omega) \,$\\
\hline
 $135$ &  6.55 & 2.27 & .944 \\ 
 $300$  &  6.55 & 2.37 & 1.14 \\ 
\hline\hline 
\end{tabular}
\end{center}
\end{table}

\end{document}